\documentclass[a4paper]{llncs}

\usepackage{amssymb}
\usepackage{graphicx}
\usepackage{amsmath}

\newcommand{\keywords}[1]{\par\addvspace\baselineskip
\noindent\keywordname\enspace\ignorespaces#1}

\begin{document}

\mainmatter  % start of an individual contribution

\title{About the Linear Complexity of Ding-Hellesth Generalized
Cyclotomic Binary Sequences of Any Period}

\titlerunning{Lecture Notes in Computer Science: Authors' Instructions}

\author{Vladimir Edemskiy}
%


\institute{Novgorod State University,\\
B. St. Petersburgskaya, 41, 173003 Velikiy Novgorod, Russia\\
\emph{Vladimir.Edemsky@novsu.ru}\\}


 \maketitle

\begin{abstract}
We defined sufficient conditions for designing Ding-Helleseth
sequences with arbitrary period  and high linear complexity for
generalized cyclotomies.  Also we discuss the method of computing
the linear complexity of Ding-Helleseth sequences in the general
case.
\end{abstract}

\section{Introduction}

For cryptographic applications, the linear complexity ($L$) of a
sequence is an important merit factor. It may be defined as the
length of the shortest linear feedback shift register that is
capable of generating the sequence. The feedback function of this
shift register can be deduced from knowledge of just $2L$
consecutive digits of the sequence. Thus, it is reasonable to
suggest that "good" sequences have $L > N / 2$ (where $N$ denotes
the period of the sequence) \cite {r1}.

 Using classical cyclotomic classes
and generalized cyclotomic classes to construct binary sequences,
which are called classical cyclotomic sequences and generalized
cyclotomic sequences respectively, is an important method for
sequence design \cite {r1}. In their paper \cite {r2} C. Ding and T.
Helleseth first introduced a new generalized cyclotomy of order 2
with respect to $p_1^{e_1}\cdots p_t^{e_t}$, which includes
classical cyclotomy as a special case and they show how to construct
binary sequences based on this new generalized cyclotomy.  There are
many works devoted to the investigation of the properties of the
Ding-Helleseth sequences. In particular, the linear complexity of
these sequences is studied in [3-11]. Most of the papers study a
special case described in sections 1-5 of \cite {r2}. In addition,
the linear complexity of the sequences based on cyclotomic classes
of higher orders was considered for specific modules ($p^n, pq$) and
for a special case [9, 12-14].

The purpose of this paper is to find Ding-Helleseth generalized
cyclotomic sequences with arbitrary periods and  high linear
complexity. In particular, we generalize the result of Tongjiang Yan
\cite{r15}. Also we discuss the computation of the linear complexity
of these sequences in the general case.

\section{Basic Definitions and Notations}

In this section let us briefly recall the definition of
Ding-Helleseth generalized cyclotomic sequences \cite {r2}.

Let $n=p_1^{e_1}\cdots p_t^{e_t}$, when $p_1,...,p_t$ be pairwise
distinct odd primes satisfying
$$\mbox{gcd} \left (p_i^{e_i-1}(p_i-1), p_j^{e_j-1}(p_j-1) \right )=2
\mbox{  for all  }
 i\neq j
$$
and  $e_1\geq 1,...,e_t\geq 1$ be integers.

 Let $Z_n$ be the ring of residue classes modulo $n$.  According to the Chinese Remainder
 Theorem
 \begin{equation}
  \label{1}
 Z_n\cong Z_{p_1^{e_1}}\times ... \times Z_{p_t^{e_t}}
 \end{equation}
 relatively to isomorphism $\varphi (x)=(x~(\mbox{mod}~p_1^{e_1}),...,
 x~(\mbox{mod}~
 p_t^{e_t}))$. Here and hereafter $x(\mbox{mod } n)$ denotes the least
 nonnegative integer that is congruent to $x$ modulo $n$.

It is well known that exists a primitive root $g_i$  modulo
$p_i^{e_t}$. Let $D_0^{(p_i^{e_i})}=\{{g_i}^{2j}| j \in Z \}$ be the
subgroup of $Z^*_{p_i^{e_i}}$, generated by $g_i^2$, and
$D_1^{(p_i^{e_i})}=g_iD_0^{(p_i^{e_i})}$, where the arithmetic is
that of $Z_{p_i^{e_i}}$, $ i=1,2,...,t$.

Let $\textbf{\emph{a}}=(a_1,...,a_t)$ be a nonzero vector from
$\left (Z_2 \right )^t$ and
$$ I_0^{(\textbf{\emph{a}},n)}= \left \{(i_1,...,i_t) \in \left (Z_2
\right )^t \left | \right. \sum _{k=1}^t i_ka_k=0 \right \}, \qquad
I_1^{(\textbf{\emph{a}},n)}=\left (Z_2 \right )^t\smallsetminus
I_0^{(\textbf{\emph{a}},n)}.$$  By definition, put \cite {r2}
$$ E_j^{(\textbf{\emph{a}},n)}=\prod _{(i_1,...,i_t) \in
I_j^{(\textbf{\emph{a}},n)}} D_{i_1}^{(p_1^{e_1})}\times ... \times
D_{i_t}^{(p_t^{e_t})} \mbox{  and
}D_j^{(\emph{\textbf{a}},n)}=\varphi^{-1} \left
(E_j^{(\textbf{\emph{a}},n)} \right ),  j=0,1.$$

From our definition it follows that \cite {r2}
\begin{equation}
 \label{2}
Z_n^*=D_0^{(\emph{\textbf{a}},n)}\cup D_1^{(\emph{\textbf{a}},n)},
\qquad D_0^{(\emph{\textbf{a}},n)}\cap
D_1^{(\emph{\textbf{a}},n)}=\varnothing.
\end{equation}
Clearly there is an element $b\in Z_n^*$ such that
$D_1^{(\emph{\textbf{a}},n)}=bD_0^{(\emph{\textbf{a}},n)}$. The
$D_0^{(\emph{\textbf{a}},n)}$   and $D_1^{(\emph{\textbf{a}},n)}$
are called generalized cyclotomic classes of order 2 with respect
 to $\emph{\textbf{a}}$ and $n$. In the following
 $D_j^{(\emph{\textbf{a}},n)}$ will denote $D_{j~(\text{mod}~2)}^{(\emph{\textbf{a}},n)}$.

 Further, by \cite {r2} we have a partition
\begin{equation}
 \label{3}
Z_n \setminus \{0\}= \bigcup _{d|n, d>1} \frac {n}{d}  Z_d^*
\end{equation}

Let $d>1$ be a positive integer and $d|n$, and the nonzero vector
$\textbf{\emph{a}}_d= (a_1^{(d)},...,a_m^{(d)})\in \left (Z_2 \right
)^m$, where $m$ is a number of different prime numbers participating
in the factorization $d$.  By (\ref{2}) and (\ref{3}) we obtain
$$ Z_n\setminus \{0\}= \bigcup _{d|n, d>1} \frac{n}{d} \left (D_0^{(\emph{\textbf{a}}_d,d)}\cup
D_1^{(\emph{\textbf{a}}_d,d)} \right ). $$

 Let $$ C_0=\bigcup _{d|n,
d>1} \frac{n}{d} D_0^{(\emph{\textbf{a}}_d,d)}  \quad \text{and}
\quad C_1=\bigcup _{d|n, d>1} \frac{n}{d}
D_1^{(\emph{\textbf{a}}_d,d)} \cup \{0\}.$$ Then $\{ C_0,C_1 \}$ is
a partition of $Z_n$, i.e. $Z_n=C_0\cup C_1$ and $C_0\cap
C_1=\varnothing.$

 In accordance with \cite {r2}, the binary sequence $s^\infty$ is then defined by
 \begin{center} $s_i=j$ if and only
if $j~(\mbox{mod}~n) \in C_j$. \end{center} The $s^\infty$ is called
Ding-Helleseth sequence,  with the most frequently discussed options
when $\textbf{\emph{a}}_d=(0,...,0,1)$.

\section{Evaluation of the Linear Complexity of  Ding-Helleseth Sequences}

In this section we find sufficient conditions for Ding-Helleseth
sequences to have high linear complexity. Define $g$ to be the
unique solution of the following set of congruences $$g\equiv
g_i~(\mbox{mod}~p_i^{e_i}),          \qquad i=1,2,...,t.$$

\smallskip
LEMMA 1. If  $ \sum_{k=1}^m a_k^{(d)}$ is an odd number, then
$$gD_j^{(\textbf{a}_d,d)}=D_{j+1}^{(\textbf{a}_d,d)}$$
for $j=0,1$ , where the arithmetic is that of  $Z_d$.
\smallskip

\smallskip
PROOF. Let $d=p_{j_1}^{l_1}\ldots p_{j_m}^{l_m}$, where $j_k \in
\{1,2,...,t \}, k=1,2,...,m$ and integers $l_k$ satisfy the set of
inequalities  $1\leq l_k\leq e_{j_k}, k=1,2,...,m$. By the
definitions of $D_j^{(\emph{\textbf{a}}_d,d)}$ and $g$ we obtain
$$\phi \left (gD_0^{(\emph{\textbf{a}}_d,d)} \right )=\prod _{(j_1,...,j_m) \in
I_0^{(\textbf{\emph{a}}_d,d)}} \left
(g_{j_1}D_{j_1}^{(p_{j_1}^{l_1})} \right )\times ... \times \left
(g_{j_m}D_{j_m}^{(p_{j_m}^{l_{m}})} \right ),$$ where
$\phi(x)=(x~(\mbox{mod}~p_{j_1}^{l_1}),...,
x~(\mbox{mod}~p_{j_m}^{l_m}))$ or
$$\phi \left (gD_0^{(\emph{\textbf{a}}_d,d)} \right )=\prod _{(j_1,...,j_m) \in
I_0^{(\textbf{\emph{a}}_d,d)}} D_{j_1+1}^{(p_{j_1}^{l_{j_1}})}
\times ... \times D_{j_m+1}^{(p_{j_m}^{l_{j_m}})}. $$

The sum $\sum_{k=1}^m (j_k+1)a_k^{(d)}=\sum_{k=1}^m a_k^{(d)}$ for
$(j_1,...,j_m) \in I_0^{(\textbf{\emph{a}}_d,d)}$,   therefore, by
the condition of Lemma $(j_1+1,...,j_m+1)  \in
I_1^{(\textbf{\emph{a}}_d,d)}$ and by (2)  we have
$$D_{j_1+1}^{(p_{j_1}^{l_{j_1}})} \times
... \times D_{j_m+1}^{(p_{j_m}^{l_{j_m}})} \in
E_1^{(\textbf{\emph{a}}_d,d)}. $$  Then
$gD_0^{(\emph{\textbf{a}}_d,d)}\subset
D_1^{(\emph{\textbf{a}}_d,d)}$,   but as their orders are equal we
obtain $gD_0^{(\emph{\textbf{a}}_d,d)}=
D_1^{(\emph{\textbf{a}}_d,d)}$.

 The assertion  $gD_1^{(\emph{\textbf{a}}_d,d)}=
D_0^{(\emph{\textbf{a}}_d,d)}$  can be proven similarly.
\smallskip

Let $\alpha$ be a primitive $n$-th root of unity in the extension of
field $GF(2)$. Then by Blahut's theorem for the linear complexity
 $L$ of the sequence $s^\infty $ we have
\begin{equation}
 \label{4}
 L=n-\left| {\{v\;\left| {S(\alpha
^v)=0,v=0,1,...,n-1 \}} \right.} \right|,
\end{equation}
where $S(x)$ is defined by
 $S(x)=\sum \limits _{i\in C_1} x^i$.

In order to investigate the values of $S(\alpha ^v)$, let us
introduce subsidiary polynomials. Let $S_A(x)=\sum \limits _{i\in A}
x^i$, where $A$ is a subset of $Z_n$. Then for any $v=1,...,n-1$ we
obtain
\begin{equation}
 \label{5}
S_{C_0}(\alpha^v)+ S_{C_1}(\alpha^v)=0.
\end{equation}

\smallskip
LEMMA 2. If  $\sum_{k=1}^m a_k^{(d)}$ is an odd number, then for any
$v=1,2,...,n-1$ we have
$$S_{\frac {n} {d} D_1^{(\textbf{a}_d,d)}} \left (\alpha^{ vg} \right )=S_{\frac {n} {d}D_0^{(\textbf{a}_d,d)}}
\left (\alpha^{v}\right ).$$
\smallskip
\smallskip
PROOF.  By Lemma 1 $gD_1^{(\emph{\textbf{a}}_d,d)}=
D_0^{(\emph{\textbf{a}}_d,d)}$
 in the ring $Z_d$, then
 $\frac {n} {d} D_1^{(\emph{\textbf{a}}_d,d)}= \frac {n} {d} gD_0^{(\emph{\textbf{a}}_d,d)}$
 in the ring $Z_n$ and the statement of Lemma follows from
the definition of auxiliary polynomial. \smallskip

 Let $ \delta =\left \{ {\begin{array}
{l}
 1,\mbox{ if } S(1)=0, \\
  0,\mbox{ otherwise.} \\
 \end{array}}\right., $
 that is $ \delta =\left \{ {\begin{array} {l}
 1,\mbox{ if } n\equiv  3~(\mbox{mod}~4), \\
  0,\mbox{ otherwise.} \\
 \end{array}} \right. , $
 because by definition of sequence $S(1)=(n+1)/2$.

\smallskip
THEOREM 1. Suppose that for any integer $d>1$ with $d|n$ sum
$\sum_{k=1}^m a_k^{(d)}$ is an odd number; then for the linear
complexity $L$ of the sequence $s^\infty$ we have $$L\geq
(n+1)/2-\delta.$$
\smallskip
 \smallskip
PROOF.  By  definition of the sequence $s^\infty$ for all
$v=1,...,n-1$ we obtain
$$S(\alpha^v)=\sum_{d|n, d>1} S_{\frac {n} {d} D_1^{(\textbf{\emph{a}}_d,d)}}
\left (\alpha^{v}
 \right )+1.$$ Then, by Lemma 2
$$S(\alpha^{gv})=\sum_{d|n, d>1}
S_{\frac {n} {d} D_0^{(\textbf{\emph{a}}_d,d)}} \left (\alpha^{v}
\right )+1,$$ therefore  $S(\alpha^{gv})= S_{C_0} (\alpha^ v)+1$.
Hence, by (5)  we get $S(\alpha^v)+S(\alpha^{gv})=1$  for all
$v=1,...,n-1$.  So, the order of the set $$|\{v | S(\alpha^v)=0,
v=1,2,...,n-1\}|\leq (n-1)/2,$$ then by (\ref{4}) we have,  $L\geq
(n+1)/2-\delta$,  which was to be demonstrated. \smallskip

\smallskip
COROLLARY.
  If $2\equiv g~(\mbox{mod}~n)$ under conditions of Theorem 1, then
$L=n-\delta$.
\smallskip
 Indeed, in this case $S(\alpha^v)+S^2(\alpha^v)=1$ and
$S(\alpha^v)\neq 0$ for all $v=1,2,...,n-1$.

Let us make some more remarks on Theorem 1. When
$\textbf{\emph{a}}_d=(0,...,0,1)$, i.e. in the special case of
Ding-Helleseth
 generalized cyclotomic sequence, the condition of the Theorem 1 is automatically satisfied,
 so for this kind of sequences the evaluation of the linear complexity $L\geq (n+1)/2-\delta$ is always valid.
 It is easy to see that this is in accord with already known results about the linear complexity of the sequences of periods $p^n, pq$ [3-12].
It should be noted that not all the sequences examined in [3-12] can
be defined as $s^\infty$.

 Further, if $$\mbox{gcd} \left (p_i^{e_i-1}(p_i-1),
p_j^{e_j-1}(p_j-1) \right )=r  \mbox{  for all }
 i\neq j,$$   then like in [9, 12-14]  for $p_i^{e_i}$   we can define the generalized cyclotomic classes
  $H_k^{(p_i^{e_i})}, k=0,1,...,r-1$  of order  $r$.  Now, we suppose
$D_0^{(p_i^{e_i})}=\bigcup _{k=0}^{r/2-1}H_k^{(p_i^{e_i})}$ and
$D_1^{(p_i^{e_i})}=\bigcup _{k=r/2}^{d-1}H_k^{(p_i^{e_i})}$ \cite
{r13}.
  Then, for $\textbf{\emph{a}}_d=(0,...,0,1)$  the
evaluation of the linear complexity given in Theorem 1 is valid also
for generalized cyclotomic sequences built on new classes.  We can
prove it by just replacing the element  $g$ with $g^{r/2}$ in Lemmas
1 an 2. This is consistent with the results from [9, 12-14].

So, Theorem 1 establishes sufficient conditions for existence of
Ding-Helleseth sequences with high linear complexity.

In the next section we discuss how the linear complexity of the
generalized cyclotomic sequences could be computed in the general
case. In particular, we show that if amongst vectors
$\textbf{\emph{a}}_d$ there exists one with the even sum of
coordinates,
 then there exists  some $n$ for which the statement of Theorem 1 is not true.

\section{ About Computing of the Linear Complexity of Ding-Helleseth Sequences}

Now let us generalize the method of computing the linear complexity
of the generalized cyclotomic sequences with period $pq$ proposed in
\cite {r16}.
 By construction we see that
\begin{equation}
\label{6}
 S(\alpha^v)=\sum_{d|n, d>1} S_{\frac {n} {d}
D_1^{(\textbf{\emph{a}}_d,d)}} \left (\alpha^ v  \right )+1,
\end{equation}
 First, we examine the items of this sum.

If $d=p_{j_1}^{l_1} \cdots p_{j_m}^{l_m}$, then, by (\ref{1}) there
exist integers $b_i, i=1,...,m$ such that
$$b_1 \frac {n}{p_{j_1}^{l_1}}+...+b_m \frac
{n}{p_{j_m}^{l_m}}=\frac {n} {d},$$ where each $b_i, i=1,...,m$  is
uniquely determined  modulo $p_{j_i}^{l_i}$ and $b_i \not \equiv
0(\mbox{mod} p_{j_i})$.

Let $\beta_k=\alpha^{b_k n/{p_{j_k}^{l_k}}}, k=1,...,m$, then
$\beta_k$ is a primitive  $p_{j_k}^{l_k}$ -th root of unity in the
extension of the field $GF(2)$ and $$\alpha=\beta_1...\beta_m.$$

Generalizing Theorem 1 from \cite {r16} we obtain the following.

\smallskip
LEMMA 3. If $d=p_{j_1}^{l_1}...p_{j_m}^{l_m}$,  then for
$v=1,...,n-1$  we have
$$S_{\frac {n} {d} D_1^{(\textbf{a}_d,d)}} \left (\alpha^ v \right
)=\sum _{(i_1,...,i_m)  \in I_1^{(\textbf{a}_d,d)}} S_{i_1}^{ \left
(p_{j_1}^{l_1} \right ) } \left (\beta_{j_1}^ { \frac {n} {d}v }
\right ) \dots  S_{i_m}^{ \left (p_{j_m}^{l_m} \right )} \left
(\beta_{j_m}^ { \frac {n} {d}v } \right ),$$  where $S_{i_k}^{ \left
(p_{j_k}^{l_k} \right )} (x)=\sum_{i\in D_{i_k}^{ \left
(p_{j_k}^{l_k} \right )}}x^i. $
\smallskip

The method of computation of $S_{i_k}^{ \left (p_{j_k}^{l_k} \right
)} \left (\beta_{j_k}^v \right )$  over the values of the classical
cyclotomic sequences polynomial was proposed in [7-9].
 Thus, formula (\ref{6}), Lemma 3 and the results
from [10] allow us to compute the values of the polynomial
$S(\alpha^v)$, and consequently the linear complexity of
Ding-Helleseth  sequence if we know the kind of factorization of
$n$. By example let us look on the case when the conditions of
Theorem 1 does not hold.

Let $n=p_1p_2$ and $\textbf{\emph{a}}_{p_1p_2}=(1,1)$ then
$$I_1^{ ( \textbf{\emph{a}}_{p_1p_2},p_1p_2)}= \{ (0,1),(1,0) \} \quad  \text{and} \quad I_1^{(\emph{\textbf{a}}_{p_1},p_1)}= I_1^{(\textbf{\emph{a}}_{p_2},p_2)}= \{ (1)
\},$$  because by  definition
$\textbf{\emph{a}}_{p_1}=\textbf{\emph{a}}_{p_2}=(1)$. If
$d=p_1p_2$,  then $\beta_1=\alpha^{b_1p_2}$  and
$\beta_2=\alpha^{b_2 p_1}$,  where $b_1p_1+b_2p_2=1$ . Therefore
$\alpha^{p_1}=\beta_1^{p_2}$,
 $\alpha^{p_2}=\beta_2^{p_1}$.

Hence by Lemma 3 and (\ref{6}) for $n=p_1p_2$ and
 $\textbf{\emph{a}}_{p_1p_2}=(1,1)$ we obtain
 \begin{equation}
 \label{7}
 S(\alpha^v)=S_0^{(p_1)}(\beta_1^v)S_1^{(p_2)}(\beta_2^v)+S_1^{(p_1)}(\beta_1^v)S_0^{(p_2)}(\beta_2^v)+S_1^{(p_1)}(\beta_1^{p_2v})+S_1^{(p_2)}(\beta_2^{p_1v})+1.
 \end{equation}

 The properties of the polynomials $S_j^{(p_i)}(x), j,i=0,1$  were examined in \cite {r17}.

\smallskip
LEMMA 4. If $v\in Z_n^*$ then
$$ S(\alpha^v) =\left \{ {\begin{array} {l}
 0,\mbox{ if } p_1\equiv 3~(\mbox{mod}~4) \mbox{ and } p_2\equiv 3~(\mbox{mod}~4), \\
  1,\mbox{ otherwise.} \\
 \end{array}} \right.  $$
\smallskip
\smallskip
PROOF.  If $v\in Z_n^*$, then  from (\ref{7}), we obtain
\begin{equation}
\label{8}
 S(\alpha^v)=S_1^{(p_1)}(\beta_1^v)+S_0^{(p_2)}(\beta_2^v)+S_1^{(p_1)}(\beta_1^{p_2v})+S_1^{(p_2)}(\beta_2^{p_1v}).
 \end{equation}
 Let us consider two cases

1) If $p_1\equiv p_2\equiv 3~(\mbox{mod}~4)$,  then in accordance
with the law of quadratic reciprocity the Legendre symbols $\left (
\frac {p_1} {p_2} \right )$ and $\left ( \frac {p_2} {p_1} \right )$
are different. Without loos of generality, we can assume that $\left
( \frac {p_1} {p_2} \right )=1$ . Then
$$S_1^{(p_1)}(\beta_1^{p_2v})=S_0^{(p_1)}(\beta_1^v) \mbox{ and  }
S_1^{(p_2)}(\beta_2^{p_1v})=S_1^{(p_2)}(\beta_2^v).$$   By means of
these relations we replace the last two terms in (\ref{8}) and now
obtain $S(\alpha^v)=0$ for all $v\in Z_n^*$.

2)  If $p_1\equiv 1~(\mbox{mod}~4)$ or  $p_2\equiv 1~(\mbox{mod}~4)$
 then $\left ( \frac {p_1} {p_2} \right )=\left ( \frac {p_2} {p_1}
\right )$. In this case without loos of generality, we can assume
that
$$S_1^{(p_1)}(\beta_1^{p_2v})=S_1^{(p_1)}(\beta_1^v) \quad \text{and} \quad  S_1^{(p_2)}(\beta_2^{p_1v})=S_1^{(p_2)}(\beta_2^v)$$ \
Then by \cite {r17} and (\ref{8}) we get $S(\alpha^v)=1$ for all
$v\in Z_n^*$.
\smallskip

 Lemma 4 makes it clear that the statement of
Theorem 1 is not true for $n=p_1p_2$ and
$\textbf{\emph{a}}_{p_1p_2}=(1,1)$.

With Lemma 4 we can conclude the computation of the linear
complexity.  If $n=p_1p_2$ and $\textbf{a}_{p_1p_2}=(1,1)$, then for
the sequence $s^\infty$ we have

1. $L=p_1+p_2-1$, if $p_1\equiv 3~(\mbox{mod}~8)$ and $p_2\equiv
3~(\mbox{mod}~8)$;

2. $L=p_1+(p_2-1)/2$, if $p_1\equiv 3~(\mbox{mod}~8)$ and $p_2\equiv
7~(\mbox{mod}~8)$;

3. $L=p_2+(p_1-1)/2$, if $p_1\equiv 7~(\mbox{mod}~8)$ and $p_2\equiv
3~(\mbox{mod}~8)$;

4. $L=(p_1+p_2)/2$, if $p_1\equiv 7~(\mbox{mod}~8)$ and $p_2\equiv
7~(\mbox{mod}~8)$.

\section{Conclusions}
In the paper we defined sufficient conditions for designing
Ding-Helleseth sequences with arbitrary period  and high linear
complexity for generalized cyclotomies. In particular, we have that
the most frequently considered variant of Ding-Helleseth sequences
possesses high linear complexity for any period. Also we discuss the
method of computing the linear complexity of Ding-Helleseth
sequences in the general case.

\end{document}